# Forcing scheme in pseudopotential lattice Boltzmann model for multiphase flows

Q. Li,[1※] K. H. Luo,[1*,2] and X. J. Li[3]

[1]School of Engineering Sciences, University of Southampton, Southampton SO17 1BJ, United Kingdom

[2]Center for Combustion Energy, Key Laboratory for Thermal Science and Power Engineering of Ministry of Education, Department of Thermal Engineering, Tsinghua University, Beijing 100084, China

[3]School of Civil Engineering and Mechanics, Xiangtan University, Xiangtan 411105, China

※Qing.Li@ soton.ac.uk; *Corresponding author: K.H.Luo@soton.ac.uk

The pseudo-potential lattice Boltzmann (LB) model is a widely used multiphase model in the LB community. In this model, an interaction force, which is usually implemented via a forcing scheme, is employed to mimic the molecular interactions that cause phase segregation. The forcing scheme is therefore expected to play an important role in the pseudo-potential LB model. In this paper, we aim to address some key issues about forcing schemes in the pseudo-potential LB model. Firstly, theoretical and numerical analyses will be made for Shan-Chen's forcing scheme and the exact-difference-method (EDM) forcing scheme. The nature of these two schemes and their recovered macroscopic equations will be shown. Secondly, through a theoretical analysis, we will reveal the physics behind the phenomenon that different forcing schemes exhibit different performances in the pseudo-potential LB model. Moreover, based on the analysis, we will present an improved forcing scheme and numerically demonstrate that the improved scheme can be treated as an alternative approach for achieving thermodynamic consistency in the pseudo-potential LB model.

## I. INTRODUCTION

Multiphase flows are of great interest in natural phenomena and industrial processes, such as chemical, electronic, and power-generation industries [1]. Owing to the inherent complexity of the involved phenomena in multiphase flows, simulating the behavior of multiphase flows is very challenging. In recent years, the lattice Boltzmann (LB) method is becoming an increasing popular method for simulating multiphase flows [2-4]. Unlike conventional numerical methods, which are based on the discretization of macroscopic governing equations, the LB method is based on the mesoscopic kinetic equation for particle distribution functions [5-7]. In particular, for multiphase flows, the phase segregation can emerge naturally in the LB method as the result of particle interactions [8, 9], and therefore avoids tracking the interface between different phases, which is often required by many other numerical methods for simulating multiphase flows.

In the LB community, the first multiphase LB model was proposed by Gunstensen *et al.* [10]. Ever since, many multiphase LB models have been developed. Generally, these models can be classified into four categories: the color-gradient model [10, 11], the pseudo-potential model [8, 9, 12-16], the free-energy model [17-21], and the kinetic-theory-based model [22-24]. Among these models, the pseudo-potential LB model, which is also called Shan-Chen model, has attracted much attention. In this model, the fluid interactions are modeled by an artificial interparticle potential and the phase separation is achieved by imposing a short-range attraction between different phases. Because of its conceptual simplicity and computational efficiency, the pseudo-potential model is widely used in LB simulations of multiphase flows. However, it has also received extensive criticism on the problems of large spurious currents and thermodynamic inconsistency [24, 13]. The problem of spurious currents has been recently addressed by many researchers and some techniques that can reduce the spurious currents have been proposed, such as using higher-order isotropic discrete gradient operator [12] or midrange pseudo-potential [13].

In the pseudo-potential LB model, the interaction force is usually incorporated via a forcing scheme. Therefore, the forcing scheme is expected to play an important role and affect the numerical accuracy and the numerical stability of the model. Currently, two forcing schemes are widely used in the pseudo-potential model: one is Shan and Chen's forcing scheme [8, 9], and the other is the Exact-Difference-Method (EDM) scheme, which is proposed by Kupershtokh *et al.* [25]. Recently, Huang *et al.* [26] and Sun *et al.* [27] have numerically investigated the performances of different forcing schemes in the pseudo-potential model. Both of them found that, in terms of numerical stability, the EDM scheme is better than the Shan-Chen scheme when the non-dimensional relaxation time $\tau < 1$. However, when $\tau > 1$, the Shan-Chen scheme is better. In addition, it was found that both the two schemes give $\tau$ -dependent coexistence curves.

The above findings are interesting. However, the reasons for these findings were not clearly discussed. Particularly, the physics behind the phenomenon that different forcing schemes have different performances has not been revealed. In this paper, we aim to address these issues through theoretical and numerical study of forcing schemes in the pseudo-potential model. First, we will make theoretical analyses of the Shan-Chen and EDM schemes. The macroscopic equations recovered from these two schemes will be given and we will show that the numerical stability (against the temperature) of the Shan-Chen and EDM schemes is related to an additional term in their recovered macroscopic equations. Furthermore, a theoretical analysis will be made to reveal the physics behind the phenomenon that different forcing schemes exhibit different performances in the pseudo-potential model. Based on the analysis, we will present an improved forcing scheme for the pseudo-potential model, and will numerically demonstrate that the improved scheme can be treated as an alternative approach for achieving thermodynamic consistency in the pseudo-potential model.

The rest of the present paper is organized as follows. Section II will briefly introduce the pseudo-potential LB model. Theoretical and numerical analyses of the Shan-Chen and EDM schemes will be given in Sec. III. In Sec. IV, the physics behind the phenomenon that different forcing schemes exhibit different performances in the pseudo-potential model will be revealed, and an improved forcing scheme will be presented. Finally, a brief conclusion will be made in Sec. V.

## II. PSEUDO-POTENTIAL LB MODEL

In the LB method, the motion of a fluid is descried by a set of discrete single-particle density distribution function. With the BGK collision operator [28], the evolution equation of the density distribution function can be written as

$$f_\alpha\left(\boldsymbol{x}+\boldsymbol{e}_\alpha\delta_t, t+\delta_t\right)-f_\alpha\left(\boldsymbol{x}, t\right)=-\frac{1}{\tau}\left[f_\alpha\left(\boldsymbol{x}, t\right)-f_\alpha^{eq}\left(\boldsymbol{x}, t\right)\right]+F_\alpha , \tag{1}$$

where $f_\alpha$ is the density distribution function, $t$ is the time, $\boldsymbol{x}$ is the particle position, $\boldsymbol{e}_\alpha$ is the discrete particle velocity along the $\alpha$ th direction, $\tau$ is the non-dimensional relaxation time, $\delta_t$ is the time step, $F_\alpha$ is the forcing term, and $f_\alpha^{eq}$ is the equilibrium density distribution function, which can be given by

$$f_\alpha^{eq}=\omega_\alpha\rho\left[1+\frac{\left(\boldsymbol{e}_\alpha\cdot\boldsymbol{u}\right)}{c_s^2}+\frac{\boldsymbol{uu}:\left(\boldsymbol{e}_\alpha\boldsymbol{e}_\alpha-c_s^2\mathbf{I}\right)}{2c_s^4}\right], \tag{2}$$

where $c_s$ is the sound speed and $\omega_\alpha$ are the weights. For the two-dimensional nine-velocity (D2Q9) lattice, the weights $\omega_\alpha$ are given by $w_0=4/9$, $\omega_{1-4}=1/9$, and $\omega_{5-8}=1/36$.

In Shan and Chen's pseudo-potential LB model, the molecular interactions that cause phase segregation are modeled by an interaction force. The interaction force is calculated from an interaction potential $\psi$, which is dependent on the local fluid density. For single-component multiphase flows, the interaction force is given by [12-14]

$$\mathbf{F} = -G\psi(\boldsymbol{x})\sum_{\alpha=1}^{N} w\left(|\boldsymbol{e}_\alpha|^2\right)\psi(\boldsymbol{x}+\boldsymbol{e}_\alpha)\boldsymbol{e}_\alpha , \tag{3}$$

where $G$ is the interaction strength and $w\left(|\boldsymbol{e}_\alpha|^2\right)$ are the weights. For the case of nearest-neighbor interactions on the D2Q9 lattice, the weights $w\left(|\boldsymbol{e}_\alpha|^2\right)$ are $w(1)=1/3$ and $w(2)=1/12$. Through the Taylor expansion, the leading terms of the interaction force can be obtained as follows [14]:

$$\mathbf{F} = -Gc^2\left[\psi\nabla\psi + \frac{1}{2}c^2\psi\nabla\left(\nabla^2\psi\right) + \cdots\right], \tag{4}$$

where $c$ is the lattice constant. According to Eq. (4), the equation of state is given by

$$p = \rho c_s^2 + \frac{Gc^2}{2}\psi^2 . \tag{5}$$

For the case of nearest-neighbor interactions, the model will give the following relation [9, 14]:

$$\int_{\rho_g}^{\rho_l}\left(p_0 - \rho c_s^2 - \frac{Gc^2}{2}\psi^2\right)\frac{\psi'}{\psi}\mathrm{d}\rho = 0 , \tag{6}$$

where $\psi' = \mathrm{d}\psi/\mathrm{d}\rho$ and $p_0 = p(\rho_l) = p(\rho_g)$, in which $\rho_l$ is the density of the liquid phase and $\rho_g$ is the density of the vapor phase.

Equation (6) is usually called mechanical stability condition. Meanwhile, in the thermodynamic theory, the Maxwell construction which determines the thermodynamics consistency is built in terms of the requirement that $\int_{\rho_g}^{\rho_l}\left[p_0 - p(\rho)\right]\mathrm{d}V = 0$, where $V \propto 1/\rho$ [2, 29]. With Eq. (5), such a relation can be rewritten as follows:

$$\int_{\rho_g}^{\rho_l}\left(p_0 - \rho c_s^2 - \frac{Gc^2}{2}\psi^2\right)\frac{1}{\rho^2}\mathrm{d}\rho = 0 . \tag{7}$$

By comparing Eq. (6) with Eq. (7), it can be found that the mechanical stability solution will not agree with the thermodynamic theory unless $\psi \propto \exp(-1/\rho)$ [9, 14]. On the other hand, in order to be consistent with the equation of state in the thermodynamic theory, the potential $\psi$ should be chosen as [24, 30]

$$\psi = \sqrt{\frac{2\left(p_{\mathrm{EOS}} - \rho c_s^2\right)}{Gc^2}}\,, \tag{8}$$

where the pressure $p_{\mathrm{EOS}}$ is given by the equation of state in the thermodynamic theory. Obviously, Eq. (8) does not satisfy the relation $\psi \propto \exp(-1/\rho)$, which means that, when the equation of state is chosen as that in the thermodynamic theory, the mechanical stability solution of the pseudo-potential model will be inconsistent with the solution given by the Maxwell construction. This is the thermodynamic inconsistency of the pseudo-potential model.

In recent years, several researchers [31] proposed to adjust the equation of state by modifying the equilibrium distribution function. However, it should be noted that, when the pressure is changed via the equilibrium distribution function, the Galilean invariance can not be ensured, which can be clearly seen in the free-energy multiphase LB models [18, 19]. Some correction terms that involves the first-order derivative of the density should be added to the equilibrium distribution function, which will make the advantages of the pseudo-potential model lost.

## III. ANALYSES OF SHAN-CHEN AND EDM SCHEMES

### A. Shan-Chen and EDM schemes

In the original pseudo-potential LB model proposed by Shan and Chen, the interaction force is incorporated into the model by shifting the velocity in the equilibrium distribution function, and the evolution equation is given by

$$f_\alpha\left(\boldsymbol{x}+\boldsymbol{e}_\alpha\delta_t, t+\delta_t\right)-f_\alpha\left(\boldsymbol{x}, t\right)=-\frac{1}{\tau}\left[f_\alpha-f_\alpha^{eq}\left(\rho,\boldsymbol{u}^{eq}\right)\right]. \tag{9}$$

The shifted equilibrium velocity $\boldsymbol{u}^{eq}$ is given as follows:

$$\boldsymbol{u}^{eq}=\boldsymbol{u}+\frac{\tau\delta_t\mathbf{F}}{\rho}, \tag{10}$$

where $\boldsymbol{u}=\sum_\alpha f_\alpha\boldsymbol{e}_\alpha/\rho$. By averaging the moment before and after the collision, the actual fluid velocity can be defined as $\boldsymbol{v}=\boldsymbol{u}+\delta_t\mathbf{F}/(2\rho)$. Equation (9) together with Eq. (10) constitutes the Shan-Chen forcing scheme.

Another forcing scheme that is widely used in the pseudo-potential model is the EDM scheme, which is proposed by Kupershtokh *et al.* [25]. In this scheme, the forcing term in Eq. (1) is given as follows:

$$F_\alpha=f_\alpha^{eq}\left(\rho,\boldsymbol{u}+\Delta\boldsymbol{u}\right)-f_\alpha^{eq}\left(\rho,\boldsymbol{u}\right), \tag{11}$$

where $\boldsymbol{u}=\sum_\alpha f_\alpha\boldsymbol{e}_\alpha/\rho$ and $\Delta\boldsymbol{u}=\mathbf{F}\delta_t/\rho$. Similarly, the actual fluid velocity in the EDM scheme is also defined as $\boldsymbol{v}=\boldsymbol{u}+\delta_t\mathbf{F}/(2\rho)$.

### B. Theoretical analysis

In this section, a theoretical analysis will be made for the Shan-Chen and EDM schemes. Before doing this, we firstly introduce the general form of forcing schemes summarized by Guo *et al.*, which is given by [32]

$$F_\alpha=\omega_\alpha\delta_t\left[\frac{\boldsymbol{B}\cdot\boldsymbol{e}_\alpha}{c_s^2}+\frac{\boldsymbol{C}:\left(\boldsymbol{e}_\alpha\boldsymbol{e}_\alpha-c_s^2\mathbf{I}\right)}{2c_s^4}\right], \tag{12}$$

where $\boldsymbol{B}$ and $\boldsymbol{C}$ are functions of $\mathbf{F}$. In 2002, Guo *et al.* investigated the discrete lattice effects of some previous forcing schemes, and they found that, in order to recover the exact Navier-Stokes equations, $\boldsymbol{B}$ and $\boldsymbol{C}$ should be chosen as follows [32]:

$$\boldsymbol{B}=B_e\mathbf{F},\quad \boldsymbol{C}=C_e\left(\boldsymbol{v}\mathbf{F}+\mathbf{F}\boldsymbol{v}\right),\quad B_e=C_e=\left(1-\frac{1}{2\tau}\right). \tag{13}$$

Meanwhile, the velocity used in the equilibrium distribution function should be equal to the actual fluid velocity $\boldsymbol{v} = \boldsymbol{u} + \delta_t \mathbf{F}/(2\rho)$ ($\boldsymbol{u} = \sum_\alpha f_\alpha \boldsymbol{e}_\alpha / \rho$). Equations (12) and (13) are the so-called Guo *et al.*'s forcing scheme.

In what follows we will show that the Shan-Chen and EDM schemes can also be written in the form of Eq. (12). For the Shan-Chen scheme, the equilibrium distribution function can be rewritten as

$$\begin{aligned} f_\alpha^{eq}\left(\rho, \boldsymbol{u}^{eq}\right) &= \omega_\alpha \rho \left[ 1 + \frac{\boldsymbol{e}_\alpha \cdot \boldsymbol{u}^{eq}}{c_s^2} + \frac{\boldsymbol{u}^{eq}\boldsymbol{u}^{eq} : \left(\boldsymbol{e}_\alpha \boldsymbol{e}_\alpha - c_s^2 \mathbf{I}\right)}{2c_s^4} \right] \\ &= f_\alpha^{eq}\left(\rho, \boldsymbol{u}\right) + \omega_\alpha \rho \delta_t \left[ \frac{\boldsymbol{e}_\alpha}{c_s^2} \cdot \frac{\tau \mathbf{F}}{\rho} + \delta_t \left( \frac{\tau \mathbf{F}}{\rho} \frac{\tau \mathbf{F}}{\rho} + 2\boldsymbol{u} \frac{\tau \mathbf{F}}{\rho} \right) : \frac{\left(\boldsymbol{e}_\alpha \boldsymbol{e}_\alpha - c_s^2 \mathbf{I}\right)}{2c_s^4} \right] \\ &= f_\alpha^{eq}\left(\rho, \boldsymbol{u}\right) + \omega_\alpha \tau \delta_t \left[ \frac{\mathbf{F} \cdot \boldsymbol{e}_\alpha}{c_s^2} + \delta_t \frac{\left(\boldsymbol{v}_{\mathrm{SC}} \mathbf{F} + \mathbf{F} \boldsymbol{v}_{\mathrm{SC}}\right) : \left(\boldsymbol{e}_\alpha \boldsymbol{e}_\alpha - c_s^2 \mathbf{I}\right)}{2c_s^4} \right], \end{aligned} \tag{14}$$

where $\boldsymbol{v}_{\mathrm{SC}} = \boldsymbol{u} + \tau \delta_t \mathbf{F}/(2\rho)$. Substituting Eq. (14) into Eq. (9), the following forcing term can be obtained for the Shan-Chen scheme:

$$F_{\alpha,\mathrm{SC}} = \omega_\alpha \delta_t \left[ \frac{\mathbf{F} \cdot \boldsymbol{e}_\alpha}{c_s^2} + \frac{\left(\boldsymbol{v}_{\mathrm{SC}} \mathbf{F} + \mathbf{F} \boldsymbol{v}_{\mathrm{SC}}\right) : \left(\boldsymbol{e}_\alpha \boldsymbol{e}_\alpha - c_s^2 \mathbf{I}\right)}{2c_s^4} \right]. \tag{15}$$

The main difference between Eq. (15) and Eq. (13) is that the velocity used in the forcing term and the parameters ($B_e$ and $C_e$) are different. Similarly, the forcing term of the EDM scheme can be rewritten as follows:

$$\begin{aligned} F_{\alpha,\mathrm{EDM}} &= f_\alpha^{eq}\left(\rho, \boldsymbol{u} + \Delta \boldsymbol{u}\right) - f_\alpha^{eq}\left(\rho, \boldsymbol{u}\right) \\ &= \omega_\alpha \rho \left[ 1 + \frac{\boldsymbol{e}_\alpha \cdot \left(\boldsymbol{u} + \Delta \boldsymbol{u}\right)}{c_s^2} + \frac{\left(\boldsymbol{u} + \Delta \boldsymbol{u}\right)\left(\boldsymbol{u} + \Delta \boldsymbol{u}\right) : \left(\boldsymbol{e}_\alpha \boldsymbol{e}_\alpha - c_s^2 \mathbf{I}\right)}{2c_s^4} \right] - f_\alpha^{eq}\left(\rho, \boldsymbol{u}\right) \\ &= \omega_\alpha \rho \left[ \frac{\boldsymbol{e}_\alpha \cdot \Delta \boldsymbol{u}}{c_s^2} + \frac{\left(\Delta \boldsymbol{u} \Delta \boldsymbol{u} + 2\boldsymbol{u} \Delta \boldsymbol{u}\right) : \left(\boldsymbol{e}_\alpha \boldsymbol{e}_\alpha - c_s^2 \mathbf{I}\right)}{2c_s^4} \right] \\ &= \omega_\alpha \delta_t \left[ \frac{\mathbf{F} \cdot \boldsymbol{e}_\alpha}{c_s^2} + \frac{\left(\boldsymbol{v}_{\mathrm{EDM}} \mathbf{F} + \mathbf{F} \boldsymbol{v}_{\mathrm{EDM}}\right) : \left(\boldsymbol{e}_\alpha \boldsymbol{e}_\alpha - c_s^2 \mathbf{I}\right)}{2c_s^4} \right], \end{aligned} \tag{16}$$

where $\boldsymbol{v}_{\mathrm{EDM}} = \boldsymbol{u} + \delta_t \mathbf{F}/(2\rho)$. It can be seen that the velocity used in the EDM scheme's forcing term is the actual fluid velocity $\boldsymbol{v}$.

The above analysis demonstrates that the Shan-Chen and EDM schemes can also be written in the general form of forcing schemes. With Eqs. (15) and (16), the differences between the Shan-Chen and EDM schemes with other forcing schemes can be easily found (see Table I). For example, the differences between the EDM scheme and Guo *et al.*'s scheme can be summarized as follows: (i) the parameters $B_e$ and $C_e$ are different: in the EDM scheme $B_e = C_e = 1$, while in Guo *et al.*'s scheme $B_e = C_e = 1 - 1/2\tau$; (ii) the velocity used in the equilibrium distribution function is different: in the EDM scheme the used velocity is $\boldsymbol{u}$, but in Guo *et al.*'s scheme the used velocity is the actual fluid velocity $\boldsymbol{v} = \boldsymbol{u} + \delta_t \mathbf{F}/(2\rho)$. In addition, it can be seen that the only difference between the Shan-Chen scheme and the EDM scheme lies in that velocity used in the forcing term is different, and the two schemes will be identical when $\tau = 1$.

Through the Chapman-Enskog (C-E) analysis, the macroscopic equations recovered from different forcing schemes can be obtained [32]. For convenience, we use $\boldsymbol{u}^* = \boldsymbol{u} + m\delta_t \mathbf{F}/\rho$ to represent the velocity used in the equilibrium distribution function and use $\bar{\boldsymbol{v}}$ to denote the velocity used in the forcing term. Then the macroscopic equations recovered from different forcing schemes take the following unified form:

$$\frac{\partial \rho}{\partial t} + \nabla \cdot \left(\rho \boldsymbol{u}^*\right) = \delta_t \left(m - \frac{1}{2}\right) \nabla \cdot \mathbf{F}, \tag{17}$$

$$\begin{aligned} \frac{\partial \left(\rho \boldsymbol{u}^*\right)}{\partial t} + \nabla \cdot \left(\rho \boldsymbol{u}^* \boldsymbol{u}^*\right) = -\nabla p + \nabla \cdot \left(2\mu \boldsymbol{S}^*\right) + \mathbf{F} + \varepsilon \delta_t \left(m - \frac{1}{2}\right) \frac{\partial \mathbf{F}}{\partial t_1} \\ + \delta_t \nabla \cdot \left[\left(\tau - \frac{1}{2}\right)\left(\boldsymbol{u}^* \mathbf{F} + \mathbf{F} \boldsymbol{u}^*\right) - \tau C_e \left(\bar{\boldsymbol{v}} \mathbf{F} + \mathbf{F} \bar{\boldsymbol{v}}\right)\right], \end{aligned} \tag{18}$$

where $\varepsilon$ is the expansion parameter in the C-E analysis, $\partial_t = \varepsilon \partial_{t1} + \varepsilon^2 \partial_{t2}$, and $S_{ij}^* = \left(\partial_i u_j^* + \partial_j u_i^*\right)/2$.

For Guo *et al.*'s scheme ($m = 1/2$, $\boldsymbol{u}^* = \bar{\boldsymbol{v}} = \boldsymbol{v}$, and $C_e = 1 - 1/2\tau$), the exact Navier-Stokes equations will be recovered. For the Shan-Chen scheme ($m = 0$, $\boldsymbol{u}^* = \boldsymbol{u}$, $\bar{\boldsymbol{v}} = \boldsymbol{u} + \tau\delta_t \mathbf{F}/2\rho$, and $C_e = 1$), Eqs. (17) and (18) can be rewritten as

$$\frac{\partial \rho}{\partial t} + \nabla \cdot (\rho \boldsymbol{u}) = -\frac{\delta_t}{2} \nabla \cdot \mathbf{F} , \tag{19}$$

$$\frac{\partial (\rho \boldsymbol{u})}{\partial t} + \nabla \cdot (\rho \boldsymbol{u}\boldsymbol{u}) = -\nabla p + \nabla \cdot (2\mu \boldsymbol{S}) + \mathbf{F} - \frac{\varepsilon \delta_t}{2} \frac{\partial \mathbf{F}}{\partial t_1} + \delta_t \nabla \cdot \left[ -\frac{1}{2} (\boldsymbol{u}\mathbf{F} + \mathbf{F}\boldsymbol{u}) - \tau^2 \frac{\mathbf{FF}}{\rho} \right]. \tag{20}$$

Similarly, for the EDM scheme ( $m = 0$ , $\boldsymbol{u}^* = \boldsymbol{u}$ , $\bar{\boldsymbol{v}} = \boldsymbol{u} + \delta_t \mathbf{F} / 2\rho$ , and $C_e = 1$ ), we have

$$\frac{\partial \rho}{\partial t} + \nabla \cdot (\rho \boldsymbol{u}) = -\frac{\delta_t}{2} \nabla \cdot \mathbf{F} , \tag{21}$$

$$\frac{\partial (\rho \boldsymbol{u})}{\partial t} + \nabla \cdot (\rho \boldsymbol{u}\boldsymbol{u}) = -\nabla p + \nabla \cdot (2\mu \boldsymbol{S}) + \mathbf{F} - \frac{\varepsilon \delta_t}{2} \frac{\partial \mathbf{F}}{\partial t_1} + \delta_t \nabla \cdot \left[ -\frac{1}{2} (\boldsymbol{u}\mathbf{F} + \mathbf{F}\boldsymbol{u}) - \tau \delta_t \frac{\mathbf{FF}}{\rho} \right]. \tag{22}$$

From the above equations we can see that the macroscopic equations recovered from the Shan-Chen and EDM schemes both contain some additional terms. These additional terms will definitely affect the numerical performance of the model. Moreover, it can be seen that Eq. (20) and Eq. (22) are nearly the same except that the coefficient before the term $\nabla \cdot (\rho^{-1} \mathbf{FF})$ is different.

### C. Numerical analysis

In the above section, the Shan-Chen and EDM schemes have been theoretically analyzed. In this section, we will show that the numerical stability (against the temperature) of the Shan-Chen and EDM schemes is related to the term $\nabla \cdot (\rho^{-1} \mathbf{FF})$ in their recovered macroscopic equations. In simulations, the Carnahan-Starling (C-S) equation of state is adopted, which is given by [30]

$$p_{\mathrm{EOS}} = \rho RT \frac{1 + \eta + \eta^2 - \eta^3}{(1 - \eta)^3} - a\rho^2 , \tag{23}$$

where $a = 0.4963 R^2 T_c^2 / p_c$ and $\eta = b\rho/4$ , in which $b = 0.18727 RT_c / p_c$ . In this work, we set $a = 1$ , $b = 4$ , $R = 1$ , $c = 1$ , $\delta_t = 1$ , $T_c = 0.094$ , and $\rho_c = 0.13044$ .

The potential $\psi$ is calculated from Eq. (8). A $200 \times 200$ lattice is adopted and a circular droplet with a radius of $r_0 = 30$ is initially placed at the center of the domain with the liquid phase inside the droplet. The periodical boundary conditions are applied in the *x*- and *y*-directions. The density field is initialized as follows [26]:

$$\rho(x,y)=\frac{\rho_l+\rho_g}{2}-\frac{\rho_l-\rho_g}{2}\tanh\left[\frac{2(r-r_0)}{W}\right], \tag{24}$$

where $W$ is the initial interface width and $r=\sqrt{(x-x_0)^2+(y-y_0)^2}$, in which $(x_0,y_0)$ is the central position of the computational domain.

The lowest reduced temperature ($T_{\min}/T_c$) predicted by different forcing schemes at $\tau<1$ is presented in Fig. 1, from which we can observe that the EDM scheme's numerical stability is better than that of the Shan-Chen scheme, and Guo *et al.*'s scheme gives the worst numerical stability. In fact, in the above section we have shown that the macroscopic equations recovered from the Shan-Chen and EDM schemes are nearly identical except that the coefficient before the term $-\nabla\cdot\left(\rho^{-1}\mathbf{FF}\right)$ is different. For the Shan-Chen scheme, the coefficient is $\tau^2$, while for the EDM scheme, the coefficient is $\tau\delta_t$. On the basis of the fact that $\tau\delta_t>\tau^2$ ($\delta_t=1$) when $\tau<1$ and the finding that the EDM scheme performs better than the Shan-Chen scheme when $\tau<1$, it is believed that the term $-\nabla\cdot\left(\rho^{-1}\mathbf{FF}\right)$ is capable of enhancing the numerical stability.

To numerically illustrate the above point, a modified EDM scheme is introduced, in which the parameter $C_e$ is set to be $1/\tau$. As a result, in the recovered momentum equation, the coefficient before the term $-\nabla\cdot\left(\rho^{-1}\mathbf{FF}\right)$ will be given by $\delta_t$. Considering $\delta_t>\tau\delta_t$ when $\tau<1$, we expect that the modified EDM scheme will be more stable than the original EDM scheme. The numerical results are shown in Fig. 2. As expected, the modified EDM scheme exhibits a better performance. For example, at $\tau=0.7$, the smallest reduced temperature is lowered to 0.57 from 0.67.

## IV. IMPROVED FORCING SCHEME

In Sec. III, we have preliminarily shown that the numerical stability of the Shan-Chen and EDM schemes is related to an additional term in their recovered macroscopic equations. In this section, a theoretical analysis will be made to reveal the related physics. Later, based on the analysis, an improved forcing scheme will be presented.

### A. Theoretical analysis

By noting that $\nabla\cdot\left(\rho^{-1}\mathbf{FF}\right)$ is the divergence of the tensor $\rho^{-1}\mathbf{FF}$, it can be found that pressure tensor in the pseudo-potential model will be changed when the Shan-Chen and EDM schemes are employed. For the problem of one-dimensional flat interface, the analytical expression (up to second-order derivatives) for the normal pressure tensor is given by [14]

$$P_n = \rho c_s^2 + \frac{Gc^2}{2}\psi^2 + \frac{Gc^4}{12}\left[\alpha\left(\frac{\mathrm{d}\psi}{\mathrm{d}n}\right)^2 + \beta\psi\frac{\mathrm{d}^2\psi}{\mathrm{d}n^2}\right], \tag{25}$$

where $n$ denotes the normal direction of the interface, and $\alpha$ and $\beta$ are coefficients determined by the discrete gradient operator. For the fourth-order isotropic discrete gradient operator (the case of nearest-neighbor interactions), $\alpha$ and $\beta$ are given by $\alpha = 0$ and $\beta = 3$, respectively.

According to Eq. (25) and the requirement that at equilibrium $P_n$ should be equal to the constant static pressure in the bulk, the following mechanical stability condition will be obtained [14]:

$$\int_{\rho_g}^{\rho_l}\left(p_0 - \rho c_s^2 - \frac{Gc^2}{2}\psi^2\right)\frac{\psi'}{\psi^{1+\varepsilon}}\mathrm{d}\rho = 0, \tag{26}$$

where $\varepsilon = -2\alpha/\beta$. In some previous work [24, 26], it was stated that $\varepsilon$ is given by $\varepsilon = 1$ for the case of nearest-neighbor interactions. Shan has clarified this issue in Ref. [14] and demonstrated that $\varepsilon$ will be given by $\varepsilon = 0$ when the nearest-neighbor interactions are applied.

The above analysis is based on the assumption that no additional terms are introduced into the macroscopic equations by the forcing scheme. However, when the Shan-Chen and EDM schemes are used, an additional term will be introduced into the normal pressure tensor $P_n$. From Eq. (4), the leading part of $-\nabla\cdot\left(\rho^{-1}\mathbf{FF}\right)$ is given by

$$-\nabla\cdot\left(\frac{\mathbf{FF}}{\rho}\right)=-G^2c^4\nabla\cdot\left(\frac{\psi^2}{\rho}\nabla\psi\nabla\psi\right)+O\left(\partial^5\right). \tag{27}$$

With Eq. (27), the normal pressure tensor should be modified as follows:

$$P_n=\rho c_s^2+\frac{Gc^2}{2}\psi^2+\frac{Gc^4}{12}\left[\left(\alpha+12G\gamma\right)\left(\frac{\mathrm{d}\psi}{\mathrm{d}n}\right)^2+\beta\psi\frac{\mathrm{d}^2\psi}{\mathrm{d}n^2}\right], \tag{28}$$

where $\gamma$ is dependent on the coefficient before the term $-\nabla\cdot\left(\rho^{-1}\mathbf{FF}\right)$ and the value of $\psi^2/\rho$. According to Eq. (28), the parameter $\varepsilon$ in Eq. (26) will be given by $\varepsilon=-2\left(\alpha+12G\gamma\right)/\beta$. Note that, when the potential $\psi$ is calculated by Eq. (8), the value of $G$ will become unimportant [30], and the only requirement for $G$ is to ensure that the whole term inside the square root in Eq. (8) is positive. For the C-S equation of state adopted in the present paper, $G$ is set to be $G=-1$.

Now the phenomenon that different forcing schemes exhibit different performances in the pseudo-potential model can be explained: different forcing schemes will lead to different values of $\varepsilon$ and then the corresponding solution will be different. For example, for the case of nearest-neighbor interactions, if $\gamma=1/4$, $\varepsilon$ will be given by $\varepsilon=2$. Then the related mechanical stability condition is given by

$$\int_{\rho_g}^{\rho_l}\left(p_0-\rho c_s^2-\frac{Gc^2}{2}\psi^2\right)\frac{\psi'}{\psi^3}\mathrm{d}\rho=0. \tag{29}$$

After some standard algebra, Eq. (29) can be transformed to

$$\left(p_0-\rho c_s^2\right)\left(-\frac{1}{2\psi^2}\right)\Bigg|_{\rho_g}^{\rho_l}-\frac{Gc^2}{2}\ln\psi\Bigg|_{\rho_g}^{\rho_l}+\int_{\rho_g}^{\rho_l}c_s^2\left(-\frac{1}{2\psi^2}\right)\mathrm{d}\rho=0. \tag{30}$$

Using Eq. (30) and the equation of state in both phases $p_0=p\left(\rho_l\right)=p\left(\rho_g\right)$, the analytical mechanical stability solution ( $p_0$, $\rho_l$, and $\rho_g$ ) of the pseudo-potential model can be obtained to arbitrary precision via numerical integration.

The mechanical stability solutions of the cases $\varepsilon=1$ and $\varepsilon=2$ are plotted in Fig. 3. For comparison, the solution given by the thermodynamic consistency requirement (the Maxwell construction) is also presented. From Fig. 3 it can be observed that there are nearly no difference in $\rho_l$ between different cases and the mechanical stability solutions are in good agreement with the solution given by the Maxwell construction. However, for the vapor phase, $\rho_g$ is found to be greatly affected by $\varepsilon$: $\rho_g$ of the case $\varepsilon = 1$ significantly deviates from the results of the case $\varepsilon = 2$ and those obtained via the Maxwell construction when $T/T_c \leq 0.9$.

The analytical density ratios ($\rho_l/\rho_g$) of the cases $\varepsilon = 1$ and $\varepsilon = 2$ are depicted in Fig. 4, which can illustrate why different schemes exhibit different numerical stability against the temperature. It can be seen that the density ratio profile of the case $\varepsilon = 1$ is far sharper than the profiles of the case $\varepsilon = 2$ and the solution given by the thermodynamic consistency requirement. Particularly, from $T/T_c = 0.7$ to $T/T_c = 0.65$, the density ratio of the case $\varepsilon = 1$ rapidly increases from 128 to 4252. According to Fig. 4, it is not hard to understand that, if the highest density ratio is fixed, the numerical stability of the forcing scheme that gives $\varepsilon = 2$ will be much better than that of the scheme with $\varepsilon = 1$. Similarly, the scheme that gives $\varepsilon = 1$ will be more stable than the scheme with $\varepsilon = 0$. And this is the reason why Guo *et al.*'s scheme gives the worst numerical stability: Guo *et al.*'s scheme introduces no additional terms into the macroscopic equations and the parameter $\varepsilon$ is given by $\varepsilon = 0$ for one-dimensional flat interface in the case of nearest-neighbor interactions.

### B. Improved forcing scheme and numerical results

Based on the above analysis, we propose an improved version of Guo *et al.*'s forcing scheme [Eq. (13)] by using a modified velocity $\mathbf{v}'$ in the scheme, which leads to

$$F_\alpha = \omega_\alpha \delta_t \left(1-\frac{1}{2\tau}\right)\left[\frac{\mathbf{F}\cdot\boldsymbol{e}_\alpha}{c_s^2}+\frac{\left(\boldsymbol{v}'\mathbf{F}+\mathbf{F}\boldsymbol{v}'\right):\left(\boldsymbol{e}_\alpha\boldsymbol{e}_\alpha - c_s^2\mathbf{I}\right)}{2c_s^4}\right]$$

$$= \omega_\alpha \delta_t \left(1-\frac{1}{2\tau}\right)\left[\frac{\left(\boldsymbol{e}_\alpha - \boldsymbol{v}'\right)}{c_s^2}+\frac{\left(\boldsymbol{e}_\alpha\cdot\boldsymbol{v}'\right)}{c_s^4}\boldsymbol{e}_\alpha\right]\cdot\mathbf{F}, \tag{31}$$

The modified velocity $\boldsymbol{v}'$ is defined as $\boldsymbol{v}' = \boldsymbol{v} + \sigma\mathbf{F}/\left(\upsilon\psi^2\right)$, where $\upsilon = \left(\tau - 0.5\right)$ is the kinematic viscosity and $\sigma$ is a constant. Obviously, when $\sigma = 0$, the scheme will reduce to Guo *et al.*'s forcing scheme. According to Eqs. (17) and (18), the macroscopic equations recovered from the improved forcing scheme are given by

$$\frac{\partial\rho}{\partial t} + \nabla\cdot\left(\rho\boldsymbol{v}\right) = 0, \tag{32}$$

$$\frac{\partial\left(\rho\boldsymbol{v}\right)}{\partial t} + \nabla\cdot\left(\rho\boldsymbol{v}\boldsymbol{v}\right) = -\nabla p + \nabla\cdot\left(2\mu\boldsymbol{S}\right) + \mathbf{F} - \delta_t\nabla\cdot\left(2\sigma\frac{\mathbf{FF}}{\psi^2}\right). \tag{33}$$

It can seen that, compared with the macroscopic equations recovered from the Shan-Chen and EDM schemes, the macroscopic equations recovered from the improved scheme do not contain any other additional terms except the needed term $-\nabla\cdot\left(2\sigma\psi^{-2}\mathbf{FF}\right)$, which yields

$$-\nabla\cdot\left(2\sigma\frac{\mathbf{FF}}{\psi^2}\right) = -2G^2c^4\sigma\nabla\cdot\left(\nabla\psi\nabla\psi\right) + O\left(\partial^5\right). \tag{34}$$

Unlike Eq. (27), the term on the right-hand side of Eq. (34) is no longer dependent on $\psi^2/\rho$, which is a local quantity.

With Eq. (34), the pressure tensor of the model is now given by

$$P_{ij} = P_{ij,\text{original}} + 2G^2c^4\sigma\partial_i\psi\partial_j\psi, \tag{35}$$

where $P_{ij,\text{original}}$ is the original pressure tensor. In the pseudo-potential model, the original pressure tensor takes the following form: $P_{ij,\text{original}} = P_b\delta_{ij} + \kappa\partial_i\psi\partial_j\psi$. It is obvious that the added term in Eq. (35) will change the value of the coefficient $\kappa$ only. Hence, the nature of the pressure tensor is retained. For the one-dimensional flat interface, the parameter $\varepsilon$ in Eq. (26) is now given by $\varepsilon = -2\left(\alpha + 24G\sigma\right)/\beta$.

Actually, from Fig. 3b it can be seen that, for a given temperature, the density $\rho_g$ given by the thermodynamic consistency requirement is larger than $\rho_g$ of the case $\varepsilon = 1$ but is smaller than $\rho_g$ of the case $\varepsilon = 2$ . This indicates that there exists an $\varepsilon$ ($1 < \varepsilon < 2$) which will make the mechanical stability solution approximately identical to the solution given by the thermodynamic consistency requirement. In other words, the thermodynamic consistency can be approximately achieved by choosing an appropriate value of $\varepsilon$ in the mechanical stability condition. With the improved forcing scheme, we can easily adjust $\varepsilon$ via $\sigma$ .

Now numerical simulations are conducted. Firstly, the problem of one-dimensional flat interface is considered. In simulations, a $100 \times 100$ lattice is employed and the Carnahan-Starling equation of state is adopted ($G = -1$). The periodical boundary condition is applied in the *y*-direction and the density field is initialized as follows:

$$\rho(y) = \rho_g + \frac{\rho_l - \rho_g}{2}\left[\tanh(y_1) - \tanh(y_2)\right], \tag{36}$$

where $y_1 = 2(y - 25)/W$ and $y_2 = 2(y - 75)/W$ . The nearest-neighbor interactions are applied ($\alpha = 0$ and $\beta = 3$). The coexistence curves of the cases $\sigma = 0.0625$ ($\varepsilon = 1$) and $\sigma = 0.125$ ($\varepsilon = 2$) are shown in Fig. 5. From the figure we can see that the numerical results are in good agreement with the analytical mechanical stability solutions, which well validates the proposed forcing scheme and confirms the expression $\varepsilon = -2(\alpha + 24G\sigma)/\beta$ .

Since the density $\rho_g$ obtained via the Maxwell construction is close to the result of the case $\varepsilon = 2$ (see Fig. 3b), it is expected that the value of the $\varepsilon$ that makes the mechanical stability solution approximately identical to the solution given by the Maxwell construction will be close to 2. Correspondingly, the value of $\sigma$ will be in the region $[0.0625, 0.125]$ and close to $0.125$. Through numerical investigations with different values of $\sigma$ , we find that the results obtained with $\sigma = 0.105$ ($\varepsilon = 1.68$) fit well with the solution of the Maxwell construction. The coexistence curves of the cases $\tau = 0.6$ and $\tau = 0.8$ are shown in Fig. 6. Good agreement can be observed in the both cases.

Furthermore, numerical simulations are also conducted for the problem of two-dimensional circular droplet. The results obtained with $\sigma = 0.105$ are shown in Fig. 7, from which it can be seen that the numerical results agree well with those given by the Maxwell construction. Meanwhile, as expected, the proposed improved scheme is capable of enhancing the numerical stability. For instance, at $\tau = 0.6$, the improved scheme works well when $T/T_c \geq 0.63$, while the Shan-Chen scheme, the EDM scheme, and Guo *et al*.'s scheme ($\sigma = 0$) will be unstable when $T/T_c < 0.86$, $0.73$, and $0.87$, respectively.

In summary, an improved forcing scheme has been presented by using a modified velocity in Guo *et al*.'s forcing scheme. In the improved scheme, a constant $\sigma$ is introduced to adjust the mechanical stability condition of the pseudo-potential model. The value of $\sigma$ can be numerically determined by fitting the mechanical stability solution with the solution given by the Maxwell construction. The proposed scheme can be readily extended to the multiple-relaxation-time (MRT) pseudo-potential model [15].

## V. CONCLUSION

In this paper, some important issues about forcing schemes in the pseudo-potential model have been studied. First, the Shan-Chen and EDM forcing schemes have been theoretically analyzed. It has been found that these two schemes can also be written in the general form of forcing schemes, which reveals the nature of these two schemes and enables the comparison of these two schemes with other schemes to be easy. Meanwhile, the macroscopic equations recovered from the Shan-Chen and EDM schemes have been shown, and it is found that the numerical stability of these two schemes is related to an additional term in their recovered macroscopic equations.

Furthermore, through a theoretical analysis, we have revealed the physics behind the phenomenon that different forcing schemes exhibit different performances in the pseudo-potential model: the mechanical stability condition is dependent on the used forcing scheme. Particularly, for the Shan-Chen and EDM schemes, the mechanical stability condition will depends on $\tau$, and this is the reason why these two schemes give $\tau$-dependent coexistence curves. Based on the analysis, we have presented an improved forcing scheme for the pseudo-potential model by using a modified velocity in Guo *et al.*'s forcing scheme, and have numerically demonstrated that the proposed scheme can be used to achieve thermodynamic consistency in the pseudo-potential model. Its mathematical base is that there exists a suitable $\varepsilon$ which can make the mechanical stability solution approximately identical to the solution given by the thermodynamic consistency requirement in a wide range of temperature. The theoretical analysis and the proposed forcing scheme can also be applied to other equations of state and more complex interactions, although only the C-S equation of state and the nearest-neighbor interactions are considered in the present paper.

## ACKNOWLEDGMENTS

This work was supported by the Engineering and Physical Sciences Research Council of the United Kingdom under Grant No. EP/I012605/1.

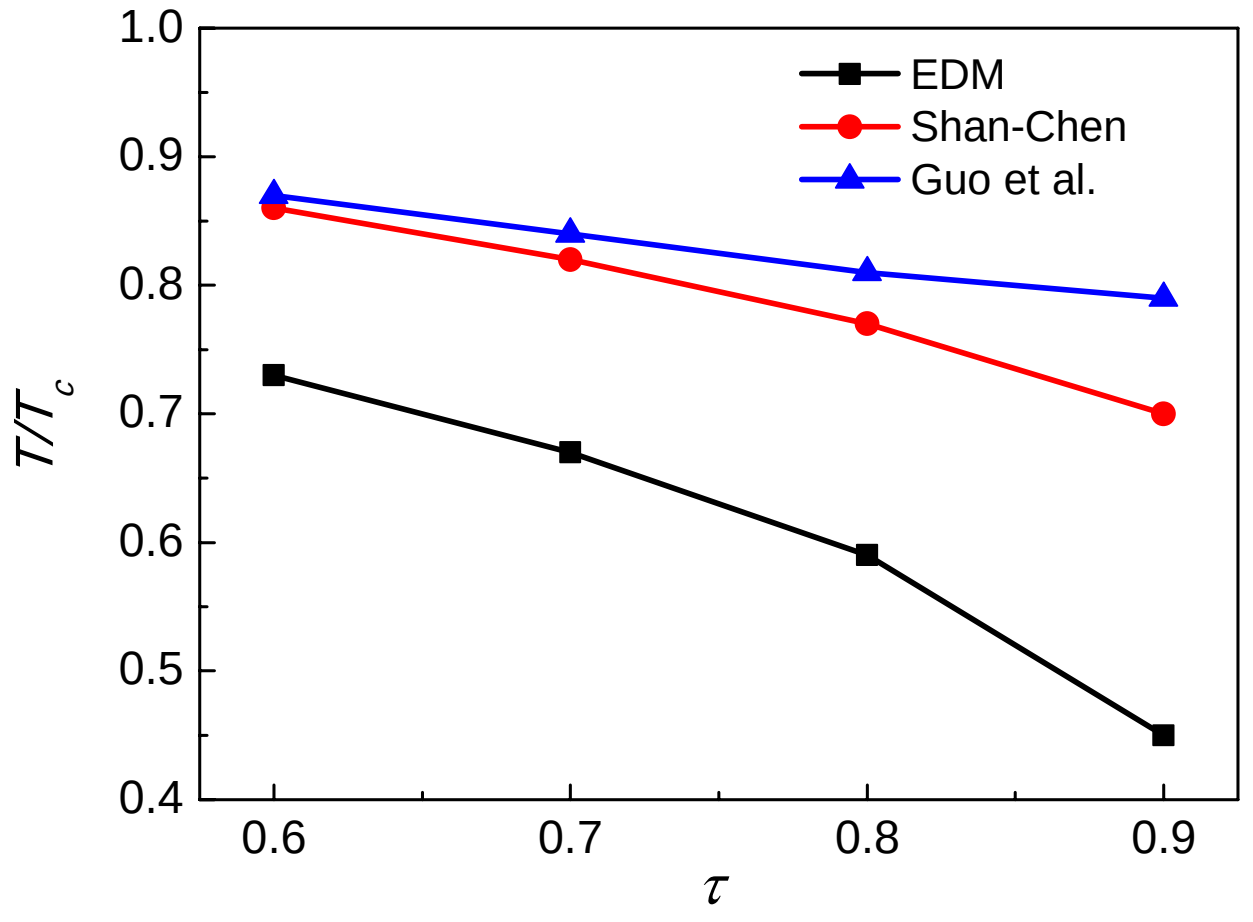

Fig. 1. Comparison of the achievable lowest temperature between the Shan-Chen scheme, the EDM scheme, and Guo *et al*.'s scheme.

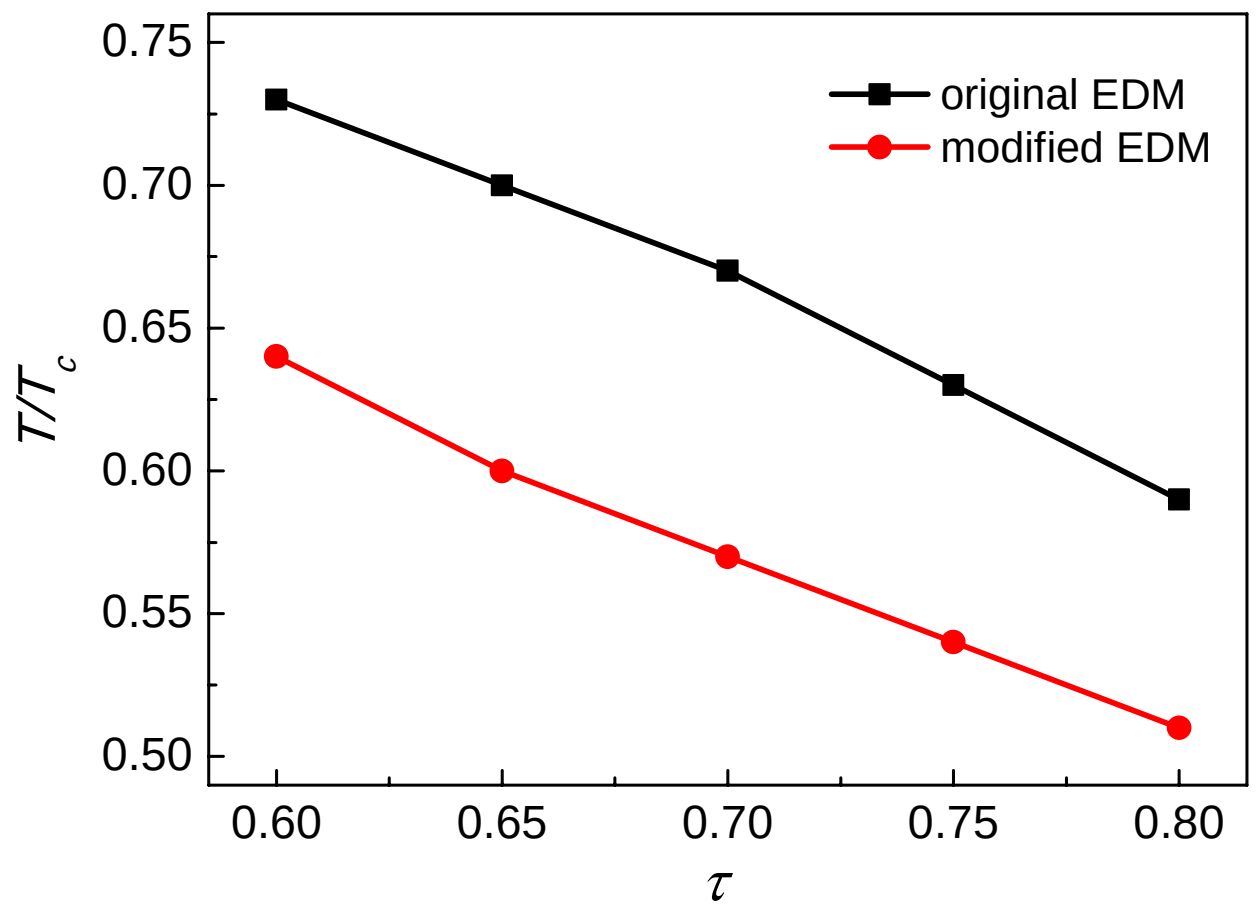

Fig. 2. Comparison of the achievable lowest temperature between the modified EDM scheme and the original EDM scheme.

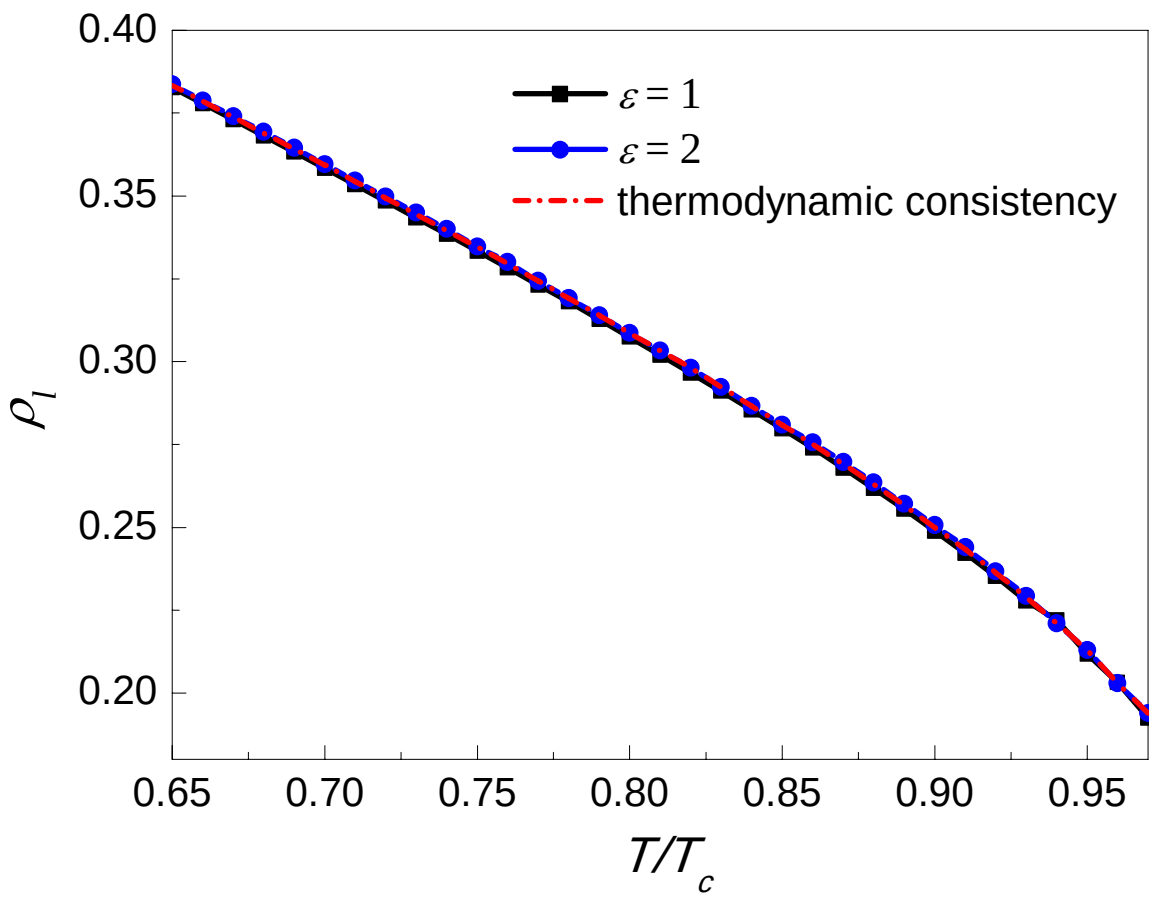

(a) density of the liquid phase

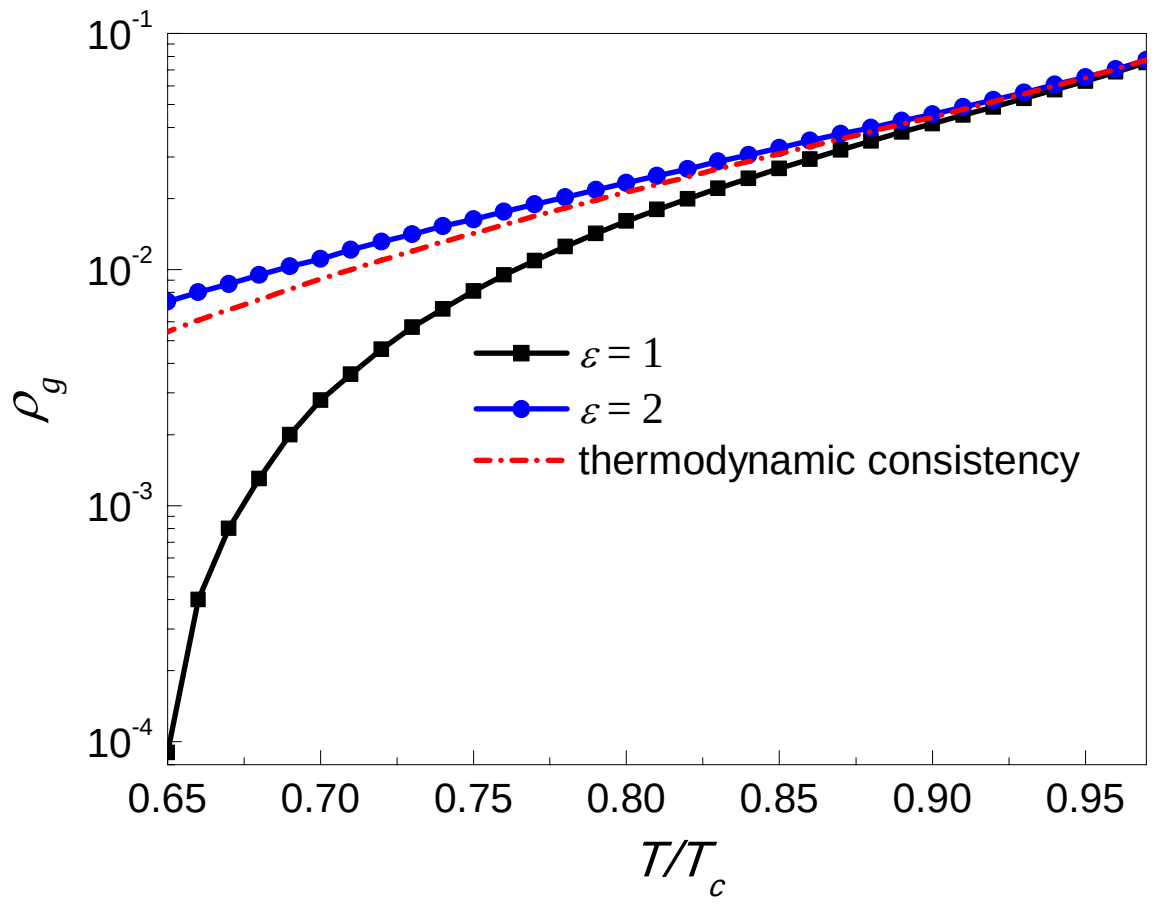

(b) density of the vapor phase

Fig. 3. Analytical mechanical stability solutions at $\varepsilon = 1$ and $\varepsilon = 2$. The dotted dash lines represent the results given by the thermodynamic consistency requirement.

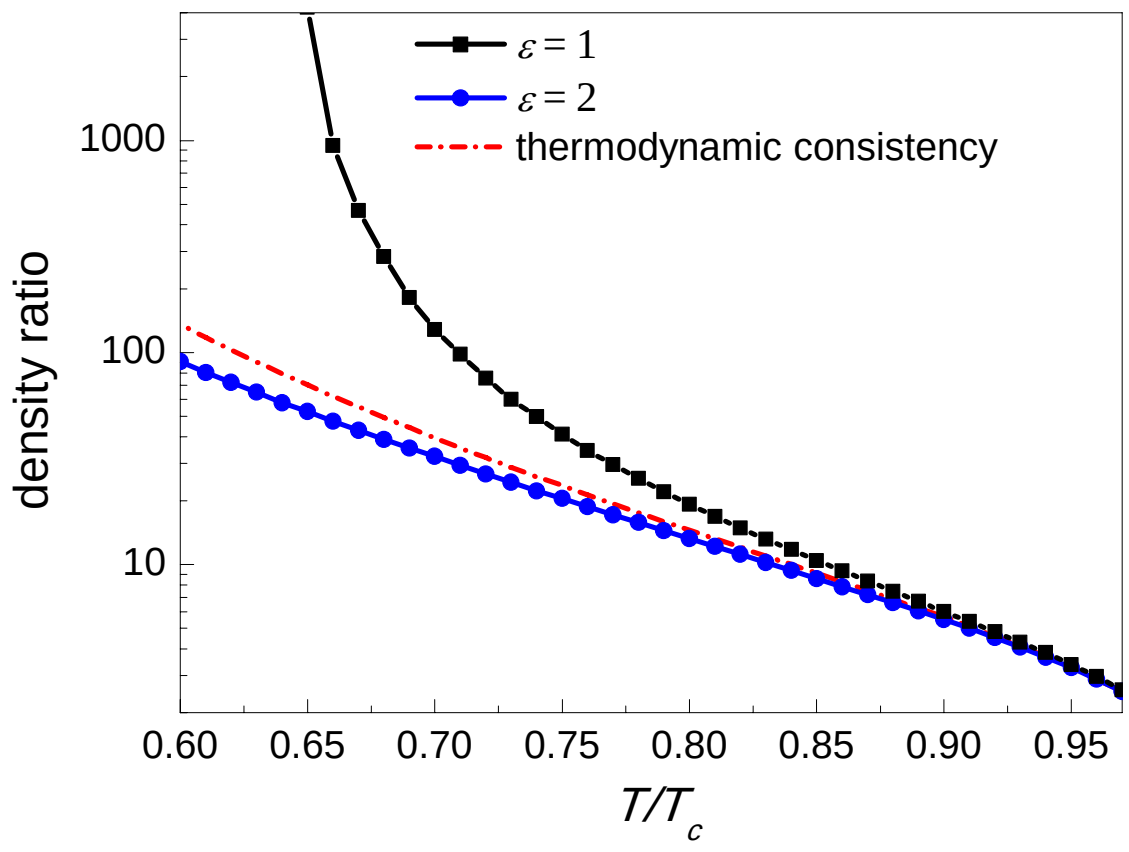

Fig. 4. The density ratios given by the analytical mechanical stability solutions at $\varepsilon = 1$ and $\varepsilon = 2$. The dotted dash line represents the results given by the thermodynamic consistency requirement.

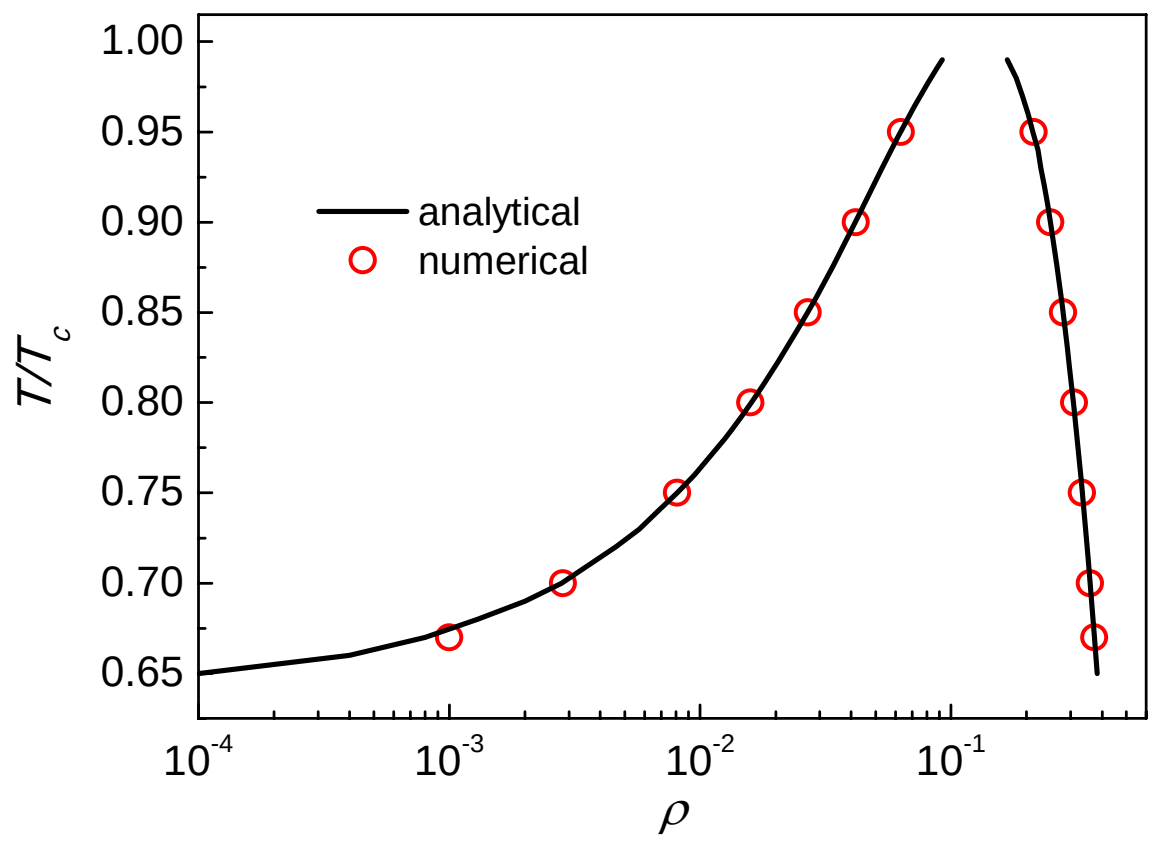

(a) $\varepsilon = 1$

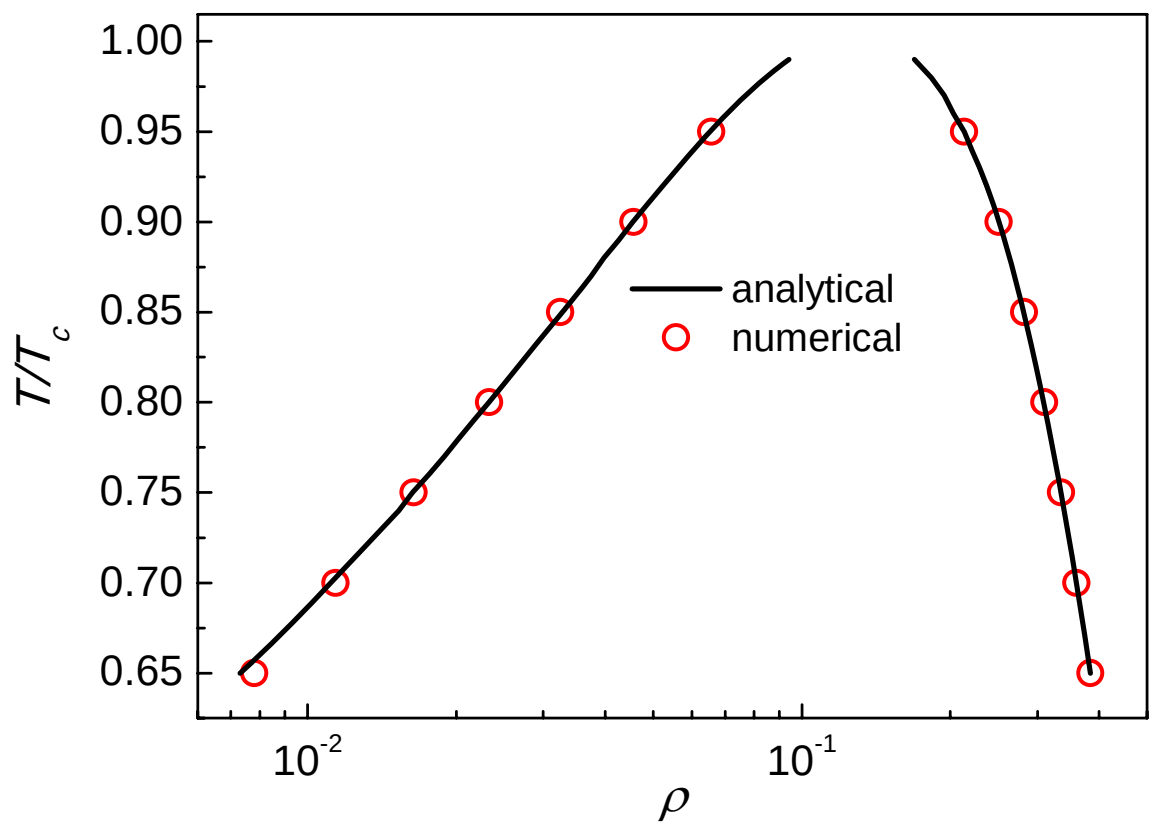

(b) $\varepsilon = 2$

Fig. 5. Simulation of one-dimensional flat interface: comparison of the numerical coexistence curves with the coexistence curves given by the analytical mechanical stability solutions.

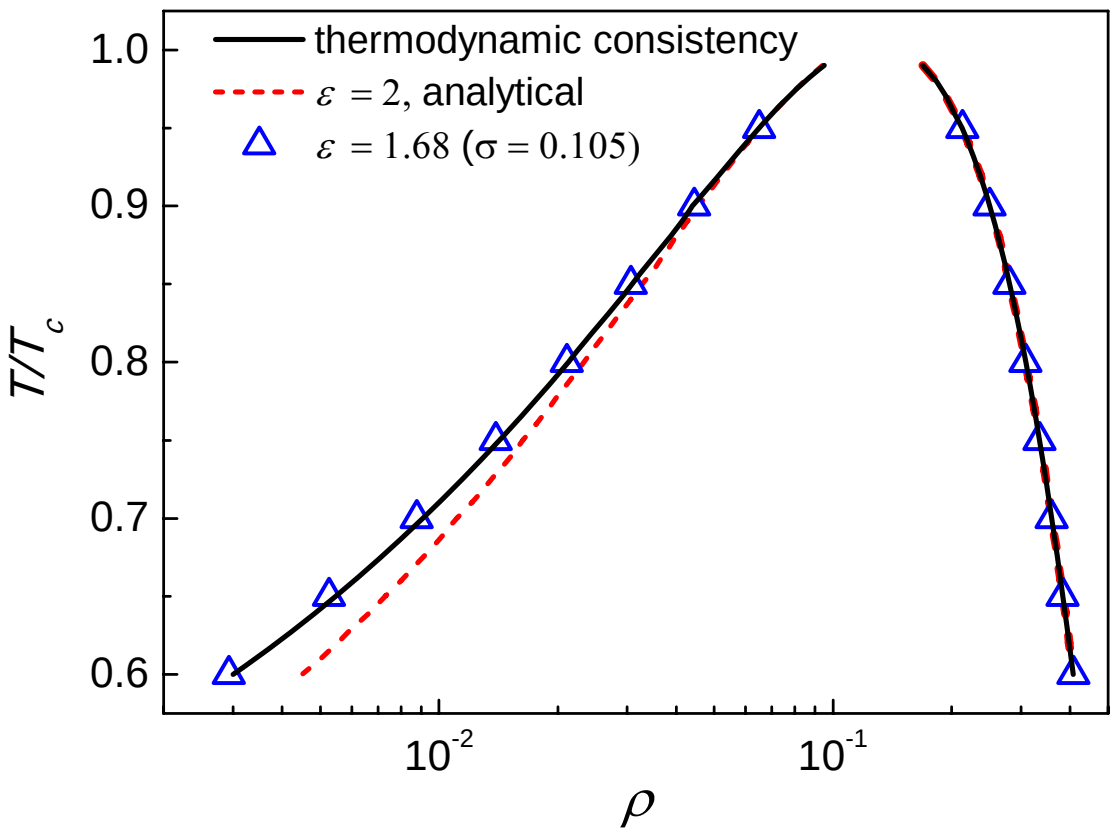

(a) $\tau = 0.6$

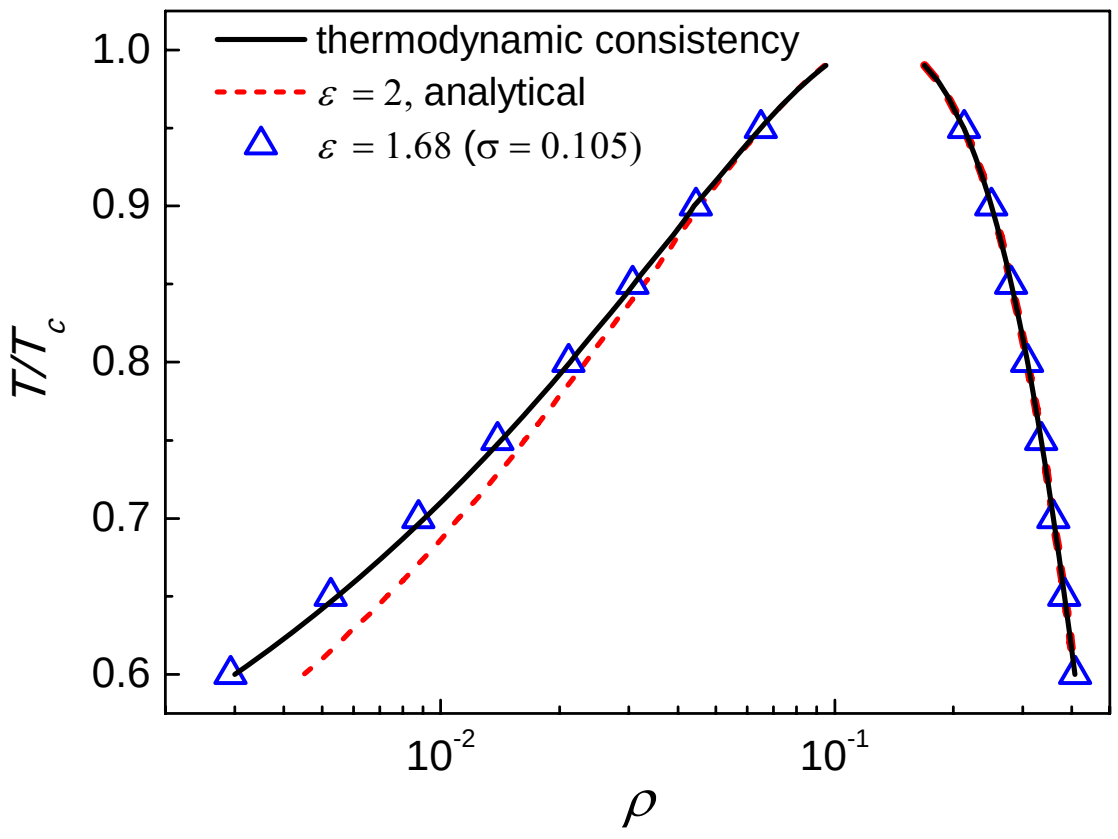

(b) $\tau = 0.8$

Fig. 6. Simulation of one-dimensional flat interface: comparison of the numerical coexistence curves obtained by $\sigma = 0.105$ with the coexistence curves given by the thermodynamic consistency requirement.

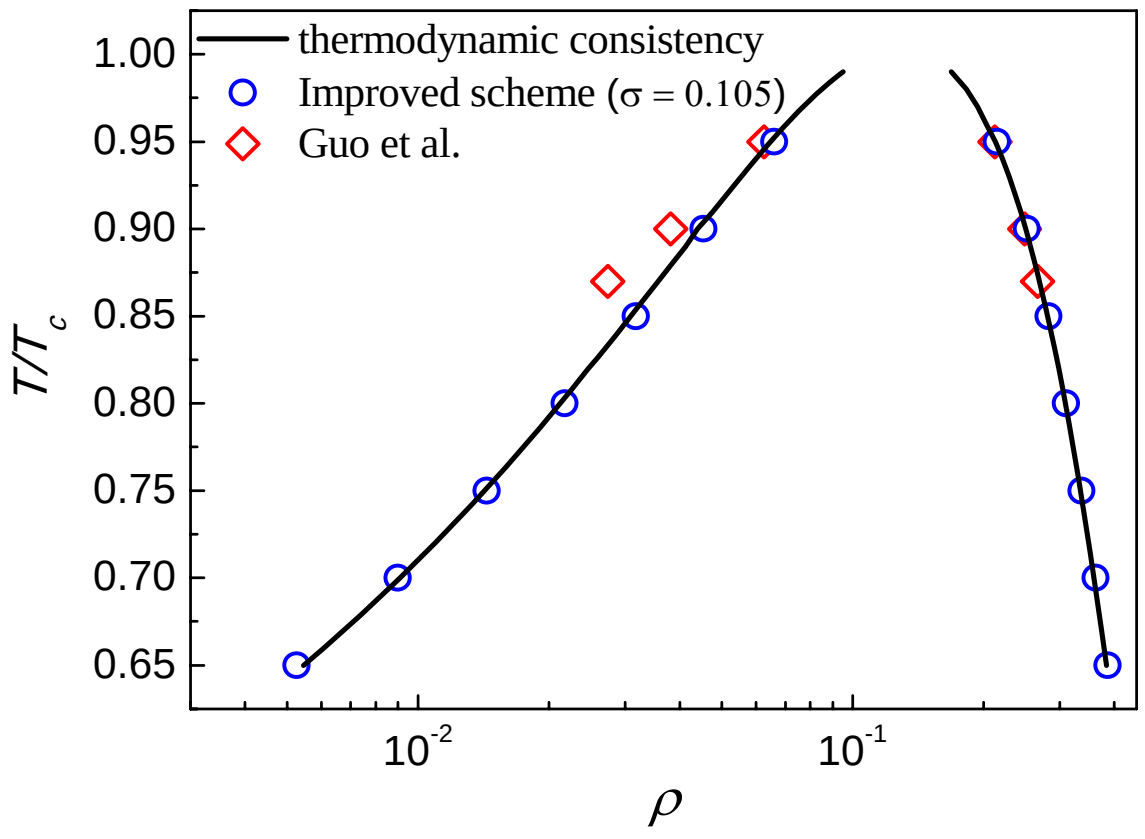

(a) $\tau = 0.6$

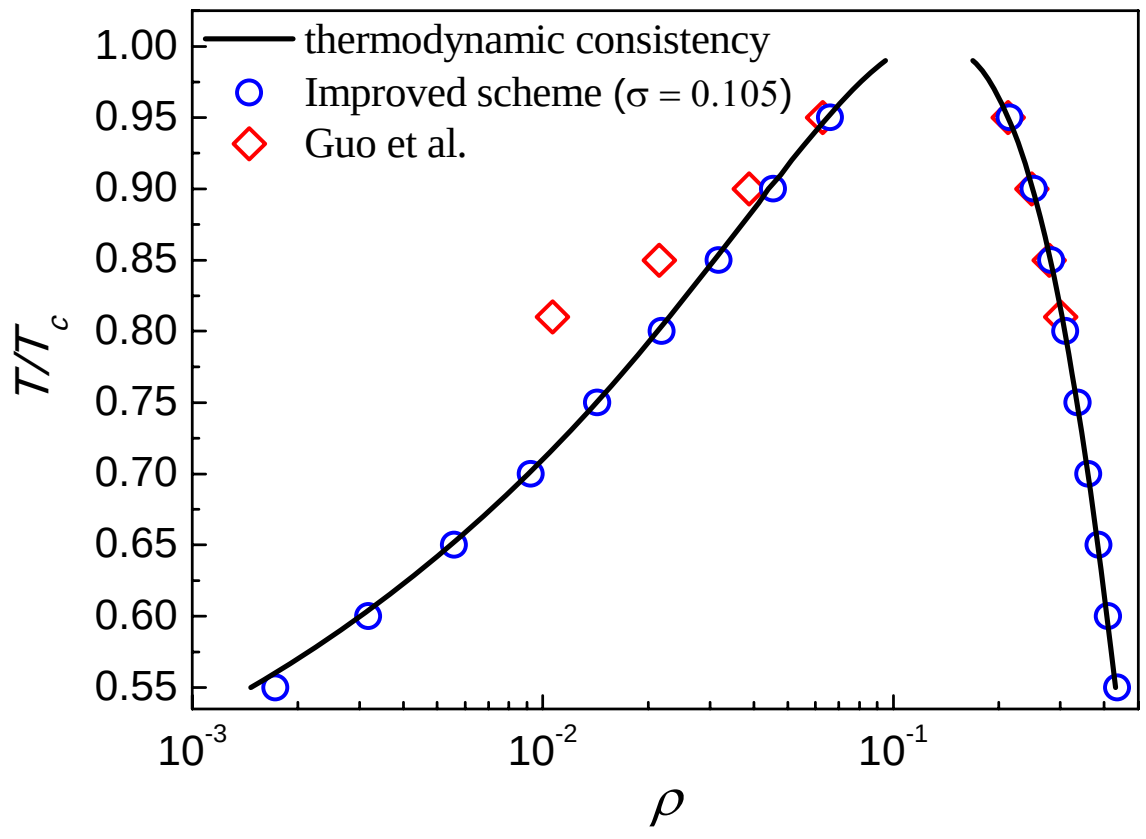

(b) $\tau = 0.8$

Fig. 7. Simulation of two-dimensional circular droplet: comparison of the numerical coexistence curves obtained by $\sigma = 0.105$, the coexistence curves obtained from Guo *et al.*'s scheme, and the coexistence curves given by the thermodynamic consistency requirement.

Table I. Comparison of different forcing schemes.

| scheme | velocity in $f_\alpha^{eq}$ | velocity in $F_\alpha$ | actual fluid velocity | $B_e$, $C_e$ |
|---|---|---|---|---|
| Shan-Chen | $\boldsymbol{u}$ | $\boldsymbol{u}+\frac{\tau\delta_t\mathbf{F}}{2\rho}$ | $\boldsymbol{u}+\frac{\delta_t\mathbf{F}}{2\rho}$ | 1 |
| EDM | $\boldsymbol{u}$ | $\boldsymbol{u}+\frac{\delta_t\mathbf{F}}{2\rho}$ | $\boldsymbol{u}+\frac{\delta_t\mathbf{F}}{2\rho}$ | 1 |
| Ladd [33] | $\boldsymbol{u}$ | $\boldsymbol{u}$ | $\boldsymbol{u}+\frac{\delta_t\mathbf{F}}{2\rho}$ | 1 |
| Guo *et al.* | $\boldsymbol{u}+\frac{\delta_t\mathbf{F}}{2\rho}$ | $\boldsymbol{u}+\frac{\delta_t\mathbf{F}}{2\rho}$ | $\boldsymbol{u}+\frac{\delta_t\mathbf{F}}{2\rho}$ | $1-\frac{1}{2\tau}$ |